\documentclass[showpacs,preprintnumbers,prd,]{revtex4}%
\usepackage{amsfonts}
\usepackage{amsmath}
\usepackage{graphicx}
\usepackage{amssymb}
\usepackage{braket}
\usepackage{float}%
\providecommand{\U}[1]{\protect\rule{.1in}{.1in}}
\begin{document}
\title{Quantum supersymmetric FRW cosmology with a scalar field}
\author{C. Ram\'{\i}rez}
\email{cramirez@fcfm.buap.mx}
\affiliation{Benem\'erita Universidad Aut\'onoma de Puebla, Facultad de Ciencias
F\'{\i}sico Matem\'aticas, P.O. Box 165, 72000 Puebla, M\'exico.}
\author{V. V\'azquez-B\'aez}
\email{manuel.vazquez@correo.buap.mx}
\affiliation{Benem\'erita Universidad Aut\'onoma de Puebla, Facultad de Ingenier\'{\i}a, 72000 Puebla, M\'exico.}

\begin{abstract}
We analyze the quantum supersymmetric cosmological FRW model with a scalar field, with a conditional probability density and the scalar field identified as time. The Hilbert space has a spinorial structure and there is only one consistent solution, with a conserved probability density. The dynamics of the scale factor is obtained from its mean value. The uncertainty relations are fulfilled and the corresponding fluctuations are consistent with a semiclassical Universe. We give two examples which turn out to have negative potential.

\end{abstract}

\maketitle

\section{Introduction}
\label{intro}
Supersymmetric quantum cosmology has been broadly studied in in the past years, see e.g. \cite{eath,moniz}. 
For uniform spaces, it can be obtained from the ``minisuperspace"
formulation \cite{ryan} of four dimensional supergravity \cite{obregon0}. The Wheeler-deWitt equation is traced back to the ``square
root'' of the hamiltonian constraint, i.e. the supersymmetric generator
constraints. Additional to these constraints, in these formulations there are also Lorentz
constraints \cite{ryan,obregon0}, which
strongly restrict the solutions \cite{obregon1}. An alternative
formulation for supersymmetric models has been given in \cite{superfield}, by means of a
`worldline' superfield approach, where the time variable is
extended to a (supersymmetry) superspace. This formulation has been worked out
for all Bianchi models \cite{bianchi}, and under the inclusion of matter \cite{superfield}. We follow an approach of
this type, by means of the covariant formulation of one-dimensional
supergravity \cite{previo}, given by the so called `new' $\Theta$ variables \cite{wess,cupa}, 
which allows to write supergravity invariant actions in a covariant Wess-Zumino gauge. 
The solution of the Wheeler-DeWitt equation is the wave function of the Universe, and depends on the degrees of freedom of the 3-space metrics. Hence, there is not an external time, as if there would not be dynamics.
This is the so called problem of time, and with all its implications is one of the main problems of quantum gravity,  subject of study since the early times of general relativity, see e.g. \cite{Kuchar,Isham,Anderson}. 
In fact, it has been argued that instead of an external time, time arises as a result of the interactions inside the Universe. 
One of the aspects of this problem is the interpretation of the wave function of the Universe and how time could arise consistently, with a conserved probability.
The operator ordering ambiguities in the constraints can be tackled considering that the Hamiltonian constraint should be self-adjoint, and that the supersymmetric generators, which classically are obtained by complex conjugation of the others, are hermitian conjugated. This requires self-adjoint bosonic momenta, and for the fermions suitable conjugation relations. This can be a problem, for instance in the FRW model,  as in the classical theory the scale factor satisfies $a\geq0$, but the hermiticity of its momentum would require the whole real line \cite{Kuchar}.
In the supersymmetric theory, the Hamiltonian constraint is modified by the addition of fermionic terms, in such a way that the anticommutator of the supersymmetry generators closes to it. Thus the Wheeler-DeWitt equation is modified, and the supersymmetric constraints give first order equations for the wave function, which is spinorial and the number of its independent components can be very high. However, the constraints can restrict this wave function strongly, leading to quite simple solutions \cite{obregon0}. Here we consider a FRW model with a real scalar field, and we argue that its supersymmetric theory could contribute to the understanding of the mentioned aspects of the problem of time. We show, for a nonvanishing superpotential, that from the solutions of the supersymmetric constraints, only one is consistent. This solution vanishes at $a=0$, allowing that its canonical conjugated momentum is self-adjoint for $a\geq0$. Further, a definition for the probability that the Universe has some value of $a$ is given, as the conditional probability that the Universe has this value of $a$ and a value $\phi$ of the scalar field, if the Universe is at $\phi$ regardless of $a$, and then the scalar field is identified with time \cite{Kuchar}. With this interpretation, the probability density depends on this time and is conserved, and the uncertainty relation of $a$ with its momentum is fulfilled. We compute then the mean value of $a$ as a function of time, and its acceleration. Considering that the actual observable regarding the scale factor is the red shift, we compute the quantum fluctuations  of the velocity of the scale factor, which reduce notably in the region corresponding to the present era, consistently with a semiclassical behavior. We give two examples, which turn out to have negative potential. The resulting universes expand from a singularity, and after one or more inflationary periods collapse to a singularity again, consistently with the results of \cite{linde2,linde1}, where this type of potentials are studied.

The outline of the paper is as follows. In the second Section we extend the FRW model with a scalar field to a worldline supergravity model and sketch the Hamiltonian analysis, in the third Section we perform the quantization, in the fourth Section we discuss the interpretation and show the results based on the examples, one of a stable potential and the other of an unstable one. In the last Section we draw some conclusions.

\section{Supersymmetric FRW model with a scalar field}

Let us consider the action of a scalar field with a four dimensional FRW metric
\begin{equation}
I=\frac{1}{\kappa^{2}}\int\left[  -3\frac{\dot{a}^{2}a}{N}+3Nka+\frac{1}{2}\frac{a^{3}\dot{\phi}^{2}}{N}-Na^{3}V\left(\phi\right)\right]  dt,
\label{actioncosmo}%
\end{equation}
where $N$ is the lapse function and $a$ is the scale factor. This Lagrangian
is invariant under time general reparametrizations. From the equations of motion of this action, with $N=1$, the Friedmann equations and the conservation equation for a perfect fluid turn out,
$\frac{\dot{a}^{2}}{a^{2}}=\rho+k$, $\frac{2\ddot{a}}{a}+\frac{\dot{a}^{2}}{a^{2}}=-(p+k)$ and 
$\dot{\rho}+\frac{3\dot{a}}{a}(p+\rho)=0$,
with $\rho=\frac{1}{2} \dot{\phi}^{2}+V(\phi)$ the energy
density, and $p=\frac{1}{2} \dot{\phi}^{2}-V(\phi) $ the pressure for the
perfect fluid $\phi(t)$. 
The Hamiltonian constraint of (\ref{actioncosmo}) is
\begin{equation}
H_{0}=-\frac{\kappa^{2}}{12a}\pi_{a}^{2}+\frac{\kappa^{2}}{2a^{3}}\pi_{\phi}^{2}-\frac{3k}{\kappa^{2}}a+\frac{1}{\kappa^{2}}a^{3}V(\phi)=0.
\label{bosonichamiltonian}%
\end{equation}
After quantization it becomes the Wheeler-DeWitt equation for the wave function of the universe, where the ordering ambiguity of the first term must be fixed.
\subsection{Supersymmetric formulation}
Supersymmetric cosmology can be obtained from one dimensional supergravity
\cite{tkachwf}, which can be formulated as general relativity on the extension
of the time coordinate to the superspace of supersymmetry \cite{cupa},
$t\rightarrow z^{M}=(t,\Theta,\bar{\Theta})$, where $\Theta$ and $\bar{\Theta
}$ are anticommuting coordinates. The basic quantities are the superfields,  see e.g. \cite{wess},
which transform as $\delta_{\zeta}\Phi(z)=-\zeta^{M}(z)\partial_{M}\Phi(z)$,
and their covariant derivatives are $\nabla_{A}\Phi={\nabla_{A}^{\ M}(z)\partial_{M}}\Phi$, 
where the index $A$ transforms as a scalar. $\nabla_{A}^{\ M}(z)$ is the superspace vielbein, 
whose superdeterminant gives the invariant density
$\mathcal{E}=\mathrm{Sdet}{\nabla_{M}}^{A}$, $\delta_{\zeta}\mathcal{E}=(-1)^{m}\partial_{M}(\zeta^{M}\mathcal{E})$, which in the covariant Wess-Zumino gauge \cite{cupa} is $\mathcal{E}=-e-\frac{i}{2}(\Theta\bar{\Psi}+\bar{\Theta}\Psi)$,
see e.g. \cite{previo}. Therefore, in order to obtain the supersymmetric cosmological model of
(\ref{actioncosmo}), superfields for the scale factor and the scalar field must be assigned
\begin{align}
\mathcal{A}\left(  t,\Theta,\bar{\Theta}\right)  =a\left(  t\right)
+i\Theta\bar{\lambda}\left(  t\right)  +i\bar{\Theta}\lambda\left(  t\right)
+B\left(  t\right)  \Theta\bar{\Theta},\\
\Phi\left(  t,\Theta,\bar{\Theta}\right)  =\phi\left(  t\right)  +i\Theta
\bar{\eta}\left(  t\right)  +i\bar{\Theta}\eta(t)+G\left(  t\right)
\Theta\bar{\Theta}. \label{superfields}%
\end{align}
The supersymmetric generalization of the action is given by
$I=I_{R}+I_{M}$ , where $I_{R}$ is the
cosmological supersymmetric generalization of the free FRW model and $I_M$ the matter term \cite{superfield,previo}
\begin{align}
I_{R}&=\frac{3}{\kappa^{2}}\int\mathcal{E}\left(  \mathcal{A}\nabla
_{\bar{\Theta}}\mathcal{A}\nabla_{\Theta}\mathcal{A}-\sqrt{k}\mathcal{A}%
^{2}\right)  d\Theta d\bar{\Theta}dt, \label{susyraction}\\
I_{M}&=\frac{1}{\kappa^{2}}\int\mathcal{E}\mathcal{A}^{3}\left[  -\frac
{1}{2}\nabla_{\bar{\Theta}}\Phi\nabla_{\Theta}\Phi+W\left(  \Phi\right)
\right]  d\Theta d\bar{\Theta}dt. \label{susymaction}%
\end{align}
Upon integration over the Grassmann parameters in (\ref{susyraction}) and (\ref{susymaction}), we find the total component
Lagrangian \cite{previo}
\begin{align}
L =\frac{1}{\kappa^2}\bigg[&-3e^{-1}a\dot{a}^{2}+3e^{-1}a\dot{a}\left(
\psi\lambda-\bar{\psi}\bar{\lambda}\right)  +\frac{1}{2}e^{-1}a^{3}\dot{\phi}^{2}-\frac{1}{2}e^{-1}a^{3}\dot{\phi}\left(  \psi\eta-\bar
{\psi}\bar{\eta}\right)  +\frac{3i}{2}a^{2}\dot{\phi}\left(
\lambda\bar{\eta}+\bar{\lambda}\eta\right)    \nonumber \\
&+3ia\left(
\lambda\dot{\bar{\lambda}}+\bar{\lambda}\dot{\lambda}\right) -\frac{i}{2}a^{3}\left(\eta\dot{\bar{\eta}}+\bar{\eta}\dot{\eta}\right) +6\sqrt{k}e\lambda\bar{\lambda}+3i\sqrt{k}a\left(\psi\lambda+\bar{\psi}\bar{\lambda}\right)
-6eaW\lambda\bar{\lambda}-\frac{3i}{2}a^{2}W\left(  \psi\lambda+\bar{\psi}\bar{\lambda}\right)  \nonumber\\
& -\frac{i}{2}a^{3}W'\left(  \psi\eta+\bar{\psi}\bar{\eta}\right) +3ea^{2}W'\left(  \bar{\lambda}\eta-\lambda\bar{\eta}\right) -ea^{3}W''\eta\bar{\eta} -\frac{3}{2}e^{-1}a\psi\bar
{\psi}\lambda\bar{\lambda}+\frac{1}{4}e^{-1}a^{3}\psi\bar{\psi}\eta
\bar{\eta} -3ea\lambda\bar{\lambda}\eta\bar{\eta} \nonumber\\
&-3eaB^{2}+6\sqrt{k}eaB-3eB\lambda\bar{\lambda}-\frac{3}{2}ea^{2}B\eta\bar{\eta}-3ea^{2}BW +\frac{1}{2}ea^{3}G^{2}+\frac{3}{2}ea^{2}G\left(
\lambda\bar{\eta}-\bar{\lambda}\eta\right)  -ea^{3}GW'\bigg].
\end{align}
The fields $B$ and $G$ do not contain kinetic terms and are eliminated solving their equations of motion
$B=\sqrt{k}-\frac{1}{2}aW-\frac{1}{2}a^{-1}\lambda\bar{\lambda}-\frac{1}{4}a\eta\bar{\eta}$ and 
$G=W'-\frac{3}{2}a^{-1}\left(  \lambda\bar{\eta}-\bar{\lambda}\eta\right)$.
Then, making the rescalings $\lambda\rightarrow\kappa a^{-1/2}\lambda$,
$\bar{\lambda}\rightarrow\kappa a^{-1/2}\bar{\lambda}$, $\eta\rightarrow\kappa
a^{-3/2}\eta$, $\bar{\eta}\rightarrow\kappa a^{-3/2}\bar{\eta}$, the Lagrangian becomes
\begin{align}
L=&-\frac{3}{\kappa^{2}}e^{-1}a\dot{a}^{2}+\frac{3}{\kappa
}e^{-1}a^{\frac{1}{2}}\dot{a}\left(  \psi\lambda-\bar{\psi}\bar{\lambda}\right)  +\frac{3k}{\kappa^{2}}ea+\frac{1}{2\kappa^{2}}e^{-1}a^{3}\dot{\phi}^{2}-\frac{1}{2\kappa}e^{-1}a^{\frac{3}{2}}\dot{\phi}\left(  \psi\eta-\bar{\psi}\bar{\eta}\right)  +\frac{3i}{2}\dot{\phi}\left(  \lambda\bar{\eta}+\bar{\lambda}\eta\right) \nonumber\\
& +3i\left(  \lambda
\dot{\bar{\lambda}}+\bar{\lambda}\dot{\lambda}\right)  -\frac{i}{2}\left(
\eta\dot{\bar{\eta}}+\bar{\eta}\dot{\eta}\right) +3\sqrt{k}ea^{-1}\lambda\bar{\lambda}-\frac{3\sqrt{k}}{2}ea^{-1}\eta\bar{\eta}+\frac{3i\sqrt{k}}{\kappa}a^{\frac{1}{2}}\left(  \psi\lambda+\bar{\psi}\bar{\lambda
}\right)  +\frac{3}{4\kappa^{2}}ea^{3}W^{2}-\frac{3\sqrt{k}}{\kappa
^{2}}ea^{2}W\nonumber\\
& -\frac{1}{2\kappa^{2}}ea^{3}W'^{2}-\frac{9}{2}eW\lambda\bar{\lambda
}+\frac{3}{4}eW\eta\bar{\eta} -\frac{3i}{2\kappa}a^{\frac{3}{2}}W\left(  \psi\lambda+\bar{\psi}\bar{\lambda}\right)  -\frac{i}{2\kappa}a^{\frac{3}{2}}W'\left(  \psi\eta+\bar{\psi
}\bar{\eta}\right)  +\frac{3}{2}eW'\left(  \bar{\lambda}\eta-\lambda
\bar{\eta}\right) \nonumber\\
& -eW''\eta\bar{\eta}-\frac{3}{2}e^{-1}\psi\bar{\psi
}\lambda\bar{\lambda} +\frac{1}{4}e^{-1}\psi\bar{\psi}\eta\bar{\eta}.\label{lagrangiana}
\end{align}
As in the bosonic case, the kinetic terms have different signs, pointing to the presence of ghosts.  
Following the usual interpretation, in this Lagrangian there are Goldstino fields as follows. Upon
substitution of the equations of motion of the auxiliary fields, the supersymetry transformations of the fermions $\lambda$ and $\eta$ become $\delta_{\zeta}\lambda=\bar\zeta\left(  \sqrt{k}-aW/2\right)  + \cdots$
and $\delta\eta=\bar\zeta W'+\cdots$. Thus if any the fields on the
r.h.s. of these equations has nonvanishing vacuum expectation value or  $k\neq0$, the
corresponding fermion is a Goldstino. However in one dimension, the appearance of this fermion does not mean necessarily that supersymmetry is spontaneously broken.
\subsection{Hamiltonian analysis}

The canonical momenta of (\ref{lagrangiana}) are $\pi_e=0$ and
\begin{align}
\pi_{a}  &  =-\frac{6}{\kappa^{2}}e^{-1}a\dot{a}+\frac{3}{\kappa}e^{-1}a^{\frac{1}{2}}\left(\psi\lambda-\bar{\psi}\bar{\lambda}\right),\label{pia}\\
\pi_{\phi}  &  =\frac{1}{\kappa^{2}}e^{-1}a^{3}\dot{\phi}-\frac{1}{2\kappa}e^{-1}a^{\frac{3}{2}}\left(\psi\eta-\bar{\psi}\bar{\eta}\right)
+\frac{3i}{2}\left(  \lambda\bar{\eta}+\bar{\lambda}\eta\right),\label{pifi}\\
\pi_{\lambda}  &  =-3i\bar{\lambda},\qquad\pi_{\bar{\lambda}}=-3i\lambda,\label{clambda}\\
\pi_{\eta}  &  =\frac{i}{2}\bar{\eta},\qquad\pi_{\bar{\eta}}=\frac{i}{2}\eta\label{ceta}.
\end{align}
Equations (\ref{clambda}) and (\ref{ceta})
are second class constraints and the corresponding Dirac brackets are
$\left\{  a,\pi_{a}\right\}  _{D} =1$, $\left\{  \phi,\pi_{\phi}\right\}_{D}=1$, 
$\left\{  \lambda,\bar{\lambda}\right\}  _{D} =-\frac{i}{6}$, $\left\{\eta,\bar{\eta}\right\}  _{D}=i$.
Using the standard definition for the Hamiltonian and solving the second class constraints, the Hamiltonian of the theory is
\begin{equation}
H=NH_{0}+\frac{1}{2}\psi S-\frac{1}{2}\bar{\psi}\bar{S}, \label{susyham}%
\end{equation}
where%
\begin{align}
H_{0}  &  =-\frac{\kappa^{2}}{12}a^{-1}\pi_{a}^{2}+\frac{\kappa^{2}}{2}a^{-3}
\pi_{\phi}^{2}-\frac{3i\kappa^{2}}{2}a^{-3}\pi_{\phi}\left(  \lambda\bar{\eta
}+\bar{\lambda}\eta\right)-\frac{3k}{\kappa^{2}}a  -\frac{3}{4\kappa^{2}}a^{3}W^{2}(\phi) +\frac{3\sqrt{k}}{\kappa^{2}}a^{2}W(\phi)+\frac{1}{2\kappa^{2}}a^{3}
W'^{2}(\phi)\nonumber\\
&  +\frac{9}{2}W(\phi)\lambda\bar{\lambda}-\frac{3}{4}W(\phi)\eta\bar{\eta}+\frac{3}{2}W'(\phi)\left(\lambda\bar{\eta}-\bar{\lambda}\eta\right)  +W''(\phi)\eta\bar{\eta} -3\sqrt{k}a^{-1}\lambda\bar{\lambda}+\frac{3\sqrt{k}}{2}a^{-1}\eta\bar{\eta} -\frac{9\kappa^{2}}{4}a^{-3}\lambda\bar{\lambda}\eta
\bar{\eta},\label{H0_SUSY}\\
S  &  =\kappa a^{-\frac{1}{2}}\pi_{a}\lambda+\kappa a^{-\frac{3}{2}}\pi_{\phi}\eta-\frac{6i\sqrt{k}}{\kappa}a^{\frac{1}{2}}\lambda+\frac{3i}{\kappa}a^{\frac{3}{2}}W(\phi)\lambda+\frac{i}{\kappa}a^{\frac{3}{2}}W'(\phi)\eta+\frac{3i\kappa
}{2}a^{-\frac{3}{2}}\lambda\eta\bar{\eta},\label{S}\\
\bar{S}  &  =\kappa a^{-\frac{1}{2}}\pi_{a}\bar\lambda+\kappa a^{-\frac{3}{2}}\pi_{\phi}\bar\eta+\frac{6i\sqrt{k}}{\kappa}a^{\frac{1}{2}}\bar\lambda-\frac{3i}{\kappa}a^{\frac{3}{2}}W(\phi)\bar\lambda-\frac{i}{\kappa}a^{\frac{3}{2}}W'(\phi)\bar\eta-\frac{3i\kappa
}{2}a^{-\frac{3}{2}}\bar\lambda\eta\bar{\eta},
\label{bar_S}%
\end{align}
which close under the Dirac brackets
\begin{equation}
\{S,\bar{S}\}_{D}=-2i H_{0},\qquad\{H_{0},S\}_{D}=\{H_{0},\bar{S}\}_{D}=0.
\label{algebra}%
\end{equation}
From (\ref{bosonichamiltonian}) we see that the scalar potential is
\begin{equation}
V(\phi)=3\sqrt{k}a^{-1}W(\phi)-\frac{3}{4}W^{2}(\phi)+\frac{1}{2}W'^{2}(\phi).\label{potencial}
\end{equation}

\section{Quantization}

Under canonical quantization, the Hamiltonian constraint is imposed as a
second order differential equation on the wave function, giving the Wheeler
DeWitt equation. In the supersymmetric case, the supercharges (\ref{S}) and
(\ref{bar_S}) give first order differential equations, from which the
Hamiltonian follows (\ref{algebra}). From the Dirac brackets we get
\begin{equation}
[a,\pi_{a}]=[\phi,\pi_{\phi}]=i,\qquad\{\lambda,\bar{\lambda}\}=\frac{1}%
{6},\qquad\{\eta,\bar{\eta}\}=-1, \label{conmutadores}%
\end{equation}
the restly (anti)commutators being zero. In particular $\lambda^{2}=\eta
^{2}=\bar{\lambda}^{2}=\bar{\eta}^{2}=0$. The bosonic momenta are represented
by derivatives and the fermionic degrees of freedom have been represented in
various ways, for instance by Dirac matrices \cite{obregon0} or as derivatives
of the canonical conjugated variables \cite{graham1}. Here we will quantize
in the simplest way, as done in \cite{graham2}, starting from a vacuum state, annihilated by the
fermionic operators $\lambda$ and $\eta$, and on which new states are created
by $\bar{\lambda}$ and $\bar{\eta}$. Then we apply the operators $S$ and
$\bar{S}$ on a general state obtained in this way, and then we look for their null
eigenstates. We use gradated Weyl ordering to fix the ordering ambiguities, i.e. products of operators which classically commute are symmetrized, and products of fermionic operators are antisymmetrized. Thus, the
quantum constraints are
\begin{align}
S  & =\frac{\kappa}{2}\left(a^{-\frac{1}{2}}\pi_a+\pi_a a^{-\frac{1}{2}}\right)\lambda+\kappa a^{-\frac{3}{2}}\pi_\phi\eta+\frac{3i}{\kappa}a^\frac{3}{2}W(\phi)\lambda+\frac{i}{\kappa}a^\frac{3}{2}W'(\phi)\eta-\frac{6i\sqrt{k}}{\kappa}a^\frac{1}{2}\lambda+\frac{3i\kappa}{4}a^{-\frac{3}{2}}\lambda[\eta,\bar{\eta}],\label{Squant}\\
\bar{S}  & =\frac{\kappa}{2}\left(a^{-\frac{1}{2}}\pi_a+\pi_a a^{-\frac{1}{2}}\right)\bar{\lambda}+\kappa a^{-\frac{3}{2}}\pi_\phi\bar\eta-\frac{3i}{\kappa}a^\frac{3}{2}W(\phi)\bar\lambda-\frac{i}{\kappa}a^\frac{3}{2}W'(\phi)\bar\eta+\frac{6i\sqrt{k}}{\kappa}a^\frac{1}{2}\bar\lambda-\frac{3i\kappa}{4}a^{-\frac{3}{2}}\bar\lambda[\eta,\bar{\eta}].\label{Sbquant}
\end{align}
The anticommutator of these operators is $\{S,\bar S\}=-2H_0$, hence the quantum hamiltonian is
\begin{align}
H_0=&-\frac{1}{12}\kappa^2\left(a^{-1}\pi_a^2+ia^{-2}\pi_a\right)+\frac{\kappa^2}{2}a^{-3}\pi_\phi^2-\frac{3i}{2}\kappa^2a^{-3}\pi_\phi(\lambda\bar\eta+\bar\lambda\eta)-\frac{3k}{\kappa^2}a-3\sqrt{k}a^{-1}[\lambda,\bar\lambda]+\frac{3\sqrt{k}}{2}a^{-1}[\eta,\bar\eta]\\
&-\frac{3}{4\kappa^2}a^3W^2(\phi)+\frac{3\sqrt{k}}{\kappa^2}a^3W(\phi)+\frac{1}{2\kappa^2}a^3W'^2(\phi)+\frac{9}{4}W(\phi)[\lambda,\bar\lambda]-\frac{3}{8}W(\phi)[\eta,\bar\eta]+\frac{3}{2}W'(\phi)(\lambda\bar\eta-\bar\lambda\eta)\\
&+\frac{1}{2}W''(\phi)[\eta,\bar\eta]-\frac{3}{2}\sqrt{k}a^{-1}[\lambda,\bar\lambda]+\frac{3\sqrt{k}}{4}a^{-1}[\eta,\bar\eta]-\frac{9\kappa^2}{16}a^{-3}[\lambda,\bar\lambda][\eta,\bar\eta]-\frac{\kappa^2}{64}a^{-3}.\label{Hquant}
\end{align}
The last term is due to the operator ordering. Thus, $\bar S=S^\dagger$, and $H_0$ is self-dual, as ensured by the second term on the right hand side of (\ref{Hquant}).

The Hilbert space is generated from the vacuum state $\ket{1}$, which
satisfies $\lambda\ket{1}=\eta\ket{1}=0$. Hence, there are
four states
\begin{equation}
\ket{1},\quad\ket{2}=\sqrt{6}\bar{\lambda}\ket{1},\quad\ket{3}=\bar{\eta
}\ket{1}\quad\mathrm{and}\quad\ket{4}=\sqrt{6}\bar{\lambda}\bar{\eta}\ket{1},
\end{equation}
which are orthogonal and have norms $\braket{2|2}=\braket{1|1}$, $\braket{3|3}=-\braket{1|1}$ and
$\braket{4|4}=-\braket{1|1}$. Therefore a general state will have the form
\begin{equation}
\ket{\Psi}=\psi_{1}(a,\phi)\ket{1}+\psi_{2}(a,\phi)\ket{2}+\psi_{3}%
(a,\phi)\ket{3}+\psi_{4}(a,\phi)\ket{4}.
\end{equation}
Hence
\begin{align}
\lambda\ket{\Psi}  &  =\frac{1}{\sqrt{6}}[\psi_{2}(a,\phi)\ket{1}+\psi
_{4}(a,\phi)\ket{3}],\\
\eta\ket{\Psi}  &  =-\psi_{3}(a,\phi)\ket{1}+\psi_{4}(a,\phi)\ket{2},\\
\bar\lambda\ket{\Psi}  &  =\frac{1}{\sqrt{6}}[\psi_{1}(a,\phi)\ket{2}+\psi
_{3}(a,\phi)\ket{4}],\\
\bar\eta\ket{\Psi}  &  =\psi_{1}(a,\phi)\ket{3}-\psi_{2}(a,\phi)\ket{4},\\
\lambda\eta\bar\eta\ket{\Psi}  &  =-\frac{1}{\sqrt{6}}\psi_{2}(a,\phi
)\ket{1},\\
\bar\lambda\eta\bar\eta\ket{\Psi}  &  =-\frac{1}{\sqrt{6}}\psi_{1}%
(a,\phi)\ket{2}.
\end{align}
Therefore, from the constraint equation $S\ket{\Psi}=0$, we get
\begin{align}
a\left(  \partial_{a}-\frac{3}{\kappa^{2}}a^{2}W+\frac{6\sqrt{k}}{\kappa^{2}%
}a+\frac{1}{8}a^{-1}\right)  \psi_{2}-\sqrt{6}\left(  \partial_{\phi}-\frac{1}%
{\kappa^{2}}a^{3}W'\right)  \psi_{3}=0,\\
\left(  \partial_{a}-\frac{3}{\kappa^{2}}a^{2}W+\frac{6\sqrt{k}}{\kappa^{2}%
}a-\frac{5}{8}a^{-1}\right)  \psi_{4}=0,\\
\left(  \partial_{\phi}-\frac{1}{\kappa^{2}}a^{3}W'\right)  \psi_{4}=0,
\end{align}
while for $\bar{S}\Psi=0$, we get
\begin{align}
\sqrt{6}\left(  \partial_{\phi}+\frac{1}{\kappa^{2}}a^{3}W'\right)  \psi
_{2}-a\left(  \partial_{a}+\frac{3}{\kappa^{2}}a^{2}W-\frac{6\sqrt{k}}%
{\kappa^{2}}a+\frac{1}{8}a^{-1}\right)  \psi_{3}=0,\\
\left(  \partial_{a}+\frac{3}{\kappa^{2}}a^{2}W-\frac{6\sqrt{k}}{\kappa^{2}%
}a-\frac{5}{8}a^{-1}\right)  \psi_{1}=0,\\
\left(  \partial_{\phi}+\frac{1}{\kappa^{2}}a^{3}W'\right)  \psi_{1}=0.
\end{align}
The equations for $\psi_{1}$ and $\psi_{4}$ can be straightforwardly solved
yielding the, up to constant factors, unique solutions
\begin{align}
\psi_{1} (a,\phi)&  = a^{\frac{5}{8}}\exp\left[  {-\frac{a^{3}W(\phi)-3\sqrt{k}a^{2}%
}{\kappa^{2}}}\right]  ,\label{psi1aT}\\
\psi_{4} (a,\phi) &  =a^{\frac{5}{8}}\exp\left[  {\frac{a^{3}W(\phi)-3\sqrt{k}a^{2}%
}{\kappa^{2}}}\right].\label{psi4aT}%
\end{align}
Further, the equations of $\psi_{2}$ and $\psi_{3}$ can be written as
\begin{align}
\left[  a\partial_{a}-\frac{3}{\kappa^{2}}a^{3}W(\phi)+\frac{6\sqrt{k}}{\kappa^{2}}a^2+\frac{1}{8}\right]\psi_{2}(a,\phi)-\sqrt{6}\left[ \partial_{\phi}-\frac{1}{\kappa^{2}}a^{3}W'(\phi)\right]  \psi_{3}(a,\phi)=0,\label{ec23-1}\\
\left[  a\partial_{a}+\frac{3}{\kappa^{2}}a^{3}W(\phi)-\frac{6\sqrt{k}}{\kappa^{2}}a^2+\frac{1}{8}\right]\psi_{3}(a,\phi)-\sqrt{6}\left[  \partial_{\phi}+\frac{1}{\kappa^{2}}a^{3}W'(\phi)\right]  \psi_{2}(a,\phi)=0.\label{ec23-2}
\end{align}
These equations might have nontrivial solutions, which could be obtained by a power series ansatz in the variable $a$, i.e. $\psi_2(a,\phi)=\sum_{n\geq0}\alpha_n(\phi)a^n$ and $\psi_3(a,\phi)=\sum_{n\geq0}\beta_n(\phi)a^n$. If there is a solution, it can be seen that the coefficients in these power series contain exponentials of the form $e^{\omega_n\phi}$ and $e^{-\omega_n\phi}$, with undetermined coefficients, where $\omega_n$ are real numerical factors. In fact, these expansions have the form
\begin{align}
\psi_2(a,\phi)&=\sum_{n\geq0}\left(c_ne^{\omega_n\phi}+d_ne^{-\omega_n\phi}\right)a^n+{\rm terms\ dependent\ on}\ e^{\pm\omega_n\phi}W,
\\
\psi_3(a,\phi)&=\sum_{n\geq0}\left(c_ne^{\omega_n\phi}-d_ne^{-\omega_n\phi}\right)a^n+{\rm terms\ dependent\ on}\ e^{\pm\omega_n\phi}W,
\end{align}
where $\omega_n=\frac{1}{\sqrt{6}}(n+\frac{1}{8})$, and $c_n$ and $d_n$ are arbitrary constants. In fact, for $W=0$, the solutions of (\ref{ec23-1}) and  (\ref{ec23-2}) are $\psi_2(a,\phi)=a^{-1/8}[f_{+}(ae^{\phi/\sqrt{6}})+f_{-}(ae^{-\phi/\sqrt{6}})]$ and  $\psi_3(a,\phi)=a^{-1/8}[f_{+}(ae^{\phi/\sqrt{6}})-f_{-}(ae^{-\phi/\sqrt{6}})]$, where $f_\pm$ are arbitrary functions, as a consequence of the fact that there are two variables.
Therefore, in general the solution to the constraint equations is
$\ket{\Psi}=C_1\,\psi_{1}(a,\phi)\ket{1}+C_2\,\psi_{2}(a,\phi)\ket{2}+C_3\,\psi_{3}(a,\phi)\ket{3}+C_4\,\psi_{4}(a,\phi)\ket{4}$,
where the factors are arbitrary constants. 
The norm of this state is
\begin{equation}
\braket{\Psi|\Psi}=\left[|C_1|^2\int|\psi_1(a,\phi)|^2dad\phi+|C_2|^2\int|\psi_2(a,\phi)|^2dad\phi-|C_3|^2\int|\psi_3(a,\phi)|^2dad\phi-|C_4|^2\int|\psi_4(a,\phi)|^2dad\phi\right]\braket{1|1}.\label{norma}
\end{equation}
If we want to have a probabilistic interpretation, it is desirable that (\ref{norma}) is well defined and always positive. Thus, we keep in this expression, as consistent solutions, only the ones whose wave functions $\psi_i(a,\phi)$ are square integrable. Moreover, the operators which represent observables must be self-adjoint. We count here as observables the phase space variables $a$, $\phi$, $\pi_a$ and $\pi_\phi$ and the Hamiltonian $H_0$.
Thus, in order that the integrals in (\ref{norma}) are well defined and the operators $a$, $\phi$, $\pi_a=-i\hbar\partial/\partial a$ and $\pi_\phi=-i\hbar\partial/\partial \phi$ are self adjoint, first their definition domains must be stated. In quantum mechanics, the integration from $-\infty$ to $\infty$ and the vanishing boundary conditions of the wave function at the limits, ensure that the momenta are self-adjoint. However, classically $a\geq0$, which poses a problem for quantization, see e.g. \cite{Isham}. Nevertheless, whatever the integration range of $a$ is, if there are nontrivial solutions for $\psi_2$ and $\psi_3$, these solutions are not square integrable, considering that $\phi$ ranges from $-\infty$ to $\infty$. Thus, we exclude these possible solutions and $\ket{\Psi}=C_1\,\psi_{1}(a,\phi)\ket{1}+C_4\,\psi_{4}(a,\phi)\ket{4}$. In this case, we observe that the solutions (\ref{psi1aT}) and (\ref{psi4aT}) vanish at $a=0$; hence $\pi_a$ and $H_0$ are self-adjoint, keeping $a\geq0$,
\begin{equation}
\int_0^\infty \overline{\psi(a,\phi)}\pi_a\psi(a,\phi)da=\int_0^\infty \overline{\pi_a\psi(a,\phi)}\psi(a,\phi)da.
\end{equation}
Further, in order that the integrals are well defined, a first condition is that the corresponding wave function vanishes as $a \to\infty$ or as $\phi\to\pm\infty$. Therefore, for $\psi_{1}$, the argument in the exponential must blow up to negative infinity, i.e. $3a^{2}\sqrt{k}-a^{3}W(\phi) \to- \infty$ as $a
\to\infty$ or as $\phi\to\pm\infty$. In particular the condition regarding $a \to\infty$,
requires that $W(\phi)$ is always positive. In the case
of $\psi_{4}$, the inverse situation holds: $W(\phi)$ must be always negative
and it should have the limit $W(\phi\to\pm\infty) \to-\infty$. Therefore, only
one of both wave functions (\ref{psi1aT}) or (\ref{psi4aT}) can be normalizable. Thus, for a given nonvanishing superpotential, there is only one square integrable wave function, whose norm (\ref{norma}) can be chosen to be positive, hence
\begin{align}
\ket{\Psi}  &  =C\psi(a,\phi)\ket{1}\equiv C\psi_1(a,\phi)\ket{1},\quad\mathrm{if}\ W(\phi)>0\quad {\rm or}\label{psi1w}\\
\ket{\Psi}  &  =C\psi(a,\phi)\ket{4}\equiv C\psi_4(a,\phi)\ket{4},\quad\mathrm{if}\ W(\phi)<0,\label{psi4w}
\end{align}
where
\begin{equation}
\psi(a,\phi)= a^{\frac{5}{8}}\exp\left[  {-\frac{a^{3}|W(\phi)|-3\sqrt{k}a^{2}}{\kappa^{2}}}\right],
\end{equation}
and $\braket{1|1}$ is chosen to be $1$ in the first case, and $-1$ in the second case.
Note that these states are bosonic. Thus, given a nonvanishing superpotential, there is only one consistent wave function.

In the following, we will restrict ourselves to the case $k=0$. If we impose the condition that these states are normalized to unity, then the normalization constant is
\begin{equation}
|C|^2=\frac{3}{\Gamma\left(  \frac{3}{4}\right)}\left\{  \int_{-\infty}^{\infty}\left[  \frac{\kappa^{2}}{2|W(\phi)|}\right]  ^{3/4}\,d\phi\right\}^{-1}. \label{C_1}%
\end{equation}
 In fact, the old problem of time shows at this stage, i.e. that from this wave function we cannot make any prediction as there is no time. In other words, if (\ref{norma}) is well defined and positive, to it corresponds rather a sort of static universe.

\section{Interpretation}

As already long ago discussed, universe has to be self contained regarding time \cite{tiempo}, i.e. the clock is part of it. In particular, it can be in the form of a scalar field \cite{Isham}. Further, nowadays the universe is classical and in fact, its gravitational evolution has been classical excluding its very first moments, when anyway the meaning of time would be expected to be blurred by the strong quantum space-time fluctuations, and in any case we would require full quantum gravity. Thus, regarding time, one would be rather interested on the classical information, which is rendered when the mean values of the observables are computed. Here we consider the FRW model with a scalar field, in a supergravity framework. Hence, the main observable is the scale factor, which on the other side is the only parameter of the space metric.  As a first step regarding interpretation, we consider the square module of the wave function as the probability density of finding a certain three geometry, i.e, in the model we are considering, a geometry with the value $a$ for the scale factor and $\phi$ for the scalar field \cite{Hartle,Kuchar}. 
We consider farther two examples, whose probability densities are shown in Figures \ref{fig_psi1_aT_1} and \ref{fig_psi1_aT_2}, and to which would correspond a frozen universe. These Figures show a well defined path along the maxima relative to the direction of the coordinate $a$, which suggests to consider evolution along this path, i.e. we would have a sort of effective wave function, given by a section of constant $\phi$, in such a way that this coordinate would be identified with time. Actually, a computation of the mean value of the scale factor gives
\begin{equation}
\braket{a}=\braket{\psi|a|\psi}=\left\vert C\right\vert^2\int_{-\infty}^\infty d\phi\int_0^\infty da\, a \left\vert \psi(a,\phi)\right\vert^2.
\end{equation}
Thus, if we restrict for a while the integration of $\phi$ to a finite interval $[\phi_1,\phi_2]$, then, from the mean value theorem we get
$\braket{a}=\int_0^\infty da\, a\,\left\vert \psi(a,\varphi)\right\vert^2/\int_0^\infty da\,\left\vert \psi(a,\varphi')\right\vert^2$,
where $\varphi,\varphi'\in[\phi_1,\phi_2]$ are such that $\int d\phi\int da\, a \vert\psi(a,\phi)\vert^2=\Delta\phi\int da\, a \vert\psi(a,\varphi)\vert^2$, and $\int d\phi\int da \vert\psi(a,\phi)\vert^2=\Delta\phi\int da \vert\psi(a,\varphi')\vert^2$. This suggests us to consider time given by the scalar field $\phi\rightarrow \tau$, with a probability amplitude \cite{Kuchar}
\begin{equation}
\Psi(a,\tau)=\left.\frac{1}{\sqrt{\int_0^\infty da\,\vert\psi(a,\phi)\vert^2}}\,\psi(a,\phi)\right\vert_{\phi=\tau}=
\sqrt{\frac{3\times 2^{3/4}}{\Gamma\left( 3/4\right)}}\left[ \frac{|W(\tau)|}{\kappa^{2}}\right]^{3/8}a^{5/8}\exp\left[  {-\frac{1}{\kappa^{2}}a^{3}|W(\tau)|}\right],
\label{onda}
\end{equation}
which is normalized at each time, $\int_0^{\infty}da\,\vert\Psi(a,\tau)\vert^2=1$.
Thus, the corresponding probability density amounts to
\begin{equation}
\left\vert\Psi(a,\tau)\right\vert^2=\left.\frac{\left\vert\psi(a,\phi)\right\vert^2}{\int_0^\infty da\,\vert\psi(a,\phi)\vert^2}\right\vert_{\phi=\tau},
\end{equation}
i.e. it is the conditional probability of the universe of being at $a$ and $\phi$, if the universe is at $\phi$ regardless of $a$.
Further, a conservation equation for this probability density can be given
$\frac{\partial}{\partial \tau}|\Psi(a,t)|^2-\frac{\partial}{\partial a}\left[\frac{1}{6}\frac{d}{d \tau}\left({\rm ln}|W(\tau)|\right)a|\Psi(a,\tau)|^2\right]=0$.
Thus, under the preceding assumption, which assigns to $\phi$ the character of clock, the classical setup arises from mean values with the probability amplitude (\ref{onda}). For instance the mean value of the scale factor is
\begin{equation}
a(\tau)=\int_0^\infty a|\Psi(a,\tau)|^2 da=\frac{\Gamma\left(13/12\right)  }{\Gamma\left(3/4\right)}\left[\frac{\kappa^{2}}{2|W(\tau)|}\right]^{1/3}. \label{a1_T}
\end{equation}
The validity of the previous assumption can be verified by a computation of the quantum fluctuations
\begin{equation}
(\Delta a)^2=\frac{\Gamma\left(3/4\right)  \Gamma\left(17/12\right)  -\Gamma\left(13/12\right)  ^{2}}%
{\Gamma\left(3/4\right)^2 }\left[\frac{\kappa^{2}}{2|W(\tau)|}\right]^{2/3}, \label{sigma1_T}%
\end{equation}
and
\begin{equation}
(\Delta\pi_a)^2=21\frac{\Gamma\left(13/12\right)  }
{\Gamma\left(3/4\right)  }\left[\frac{|W\left(\tau\right)|
}{4\kappa^{2}}\right]^{2/3},\label{SigmaPia}
\end{equation}
from which the uncertainty relation follows
\begin{equation}
\Delta a\Delta\pi_a=\frac{1}{2\Gamma\left(3/4\right)^{3/2}}\sqrt{21\,\Gamma\left(13/12\right)
\left[  \Gamma\left(3/4\right)  \Gamma\left(17/12\right)
-\Gamma\left(13/12\right)^{2}\right]  }\approx0.68.\label{incertidumbre}
\end{equation}
Note that we have set $\hbar=1$. 

In order to compute the uncertainty in the measurement of the scale factor we would be tempted to use (\ref{sigma1_T}). However, the scale factor is determined from its velocity through measurements of the red shift and the big-bang assumption. Nevertheless, the fluctuations for the velocity cannot be obtained from $d/d\tau a(\tau)$, hence we estimate them from $\dot a(t)=-6\kappa^2 a^{-1}\pi_a+$ fermionic terms, obtained from (\ref{pia}). Thus, considering that the mean values of the fermionic variables vanish, $\Delta\dot a=a^{-1}(|\dot a|\Delta a+\kappa^2/6\Delta\pi_a)$, where the quantities from the right hand side are computed from (\ref{a1_T}), (\ref{sigma1_T}) and (\ref{SigmaPia}), considering $a\rightarrow a(\tau)$. For the two examples considered, these fluctuations are shown in Figures \ref{fig_fluctuaciones_1} and \ref{fig_fluctuaciones_2}. It is remarkable that these fluctuations reduce considerably their size in the region which would correspond to the actual era, consistently with a semiclassical behavior. 

In this formulation the Hubble factor has the simple expression
\begin{equation}
H(\tau)=-\frac{13\dot W(\tau)}{12W(\tau)}, \label{Hubble_T}
\end{equation}
where $\dot{W}\equiv\frac{d}{d\tau}W$. In the following, dot and double dot will mean derivatives with respect to $\tau$.

There are some comments in order. First, for a given nonvanishing superpotential, there is only one consistent state, solution to the constraint equations. Further, this state is invariant under supersymmetry transformations, hence supersymmetry is unbroken.

In the following, we discuss two examples. These examples are somewhat representative of what can happen. The first one is of a stable potential and the second one of an unstable potential.

\subsection{Stable potential}

\begin{figure}[h]
\centering
\includegraphics[height=3.5cm,width=4.75cm]{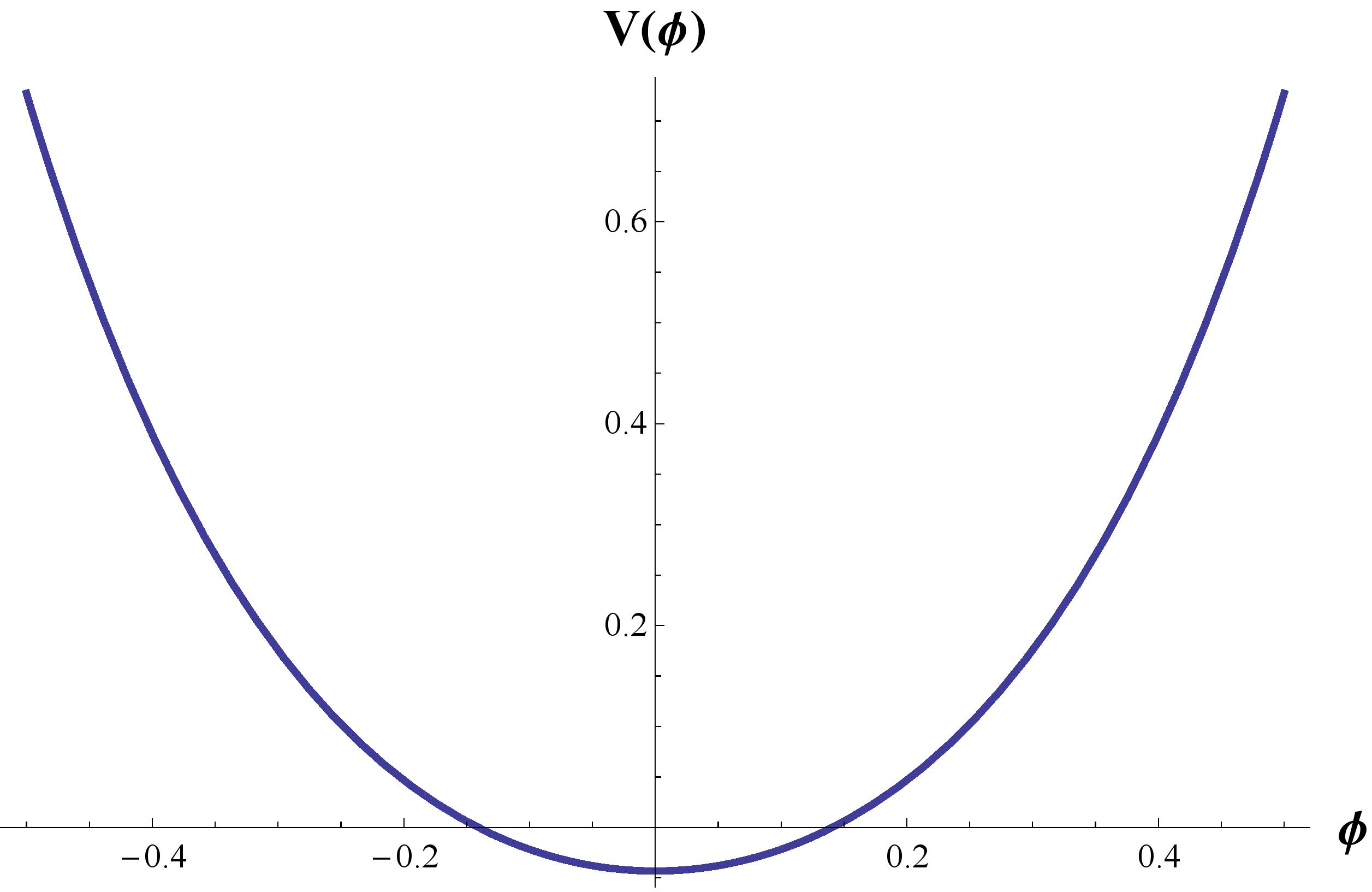}
\caption{{\protect\footnotesize {Stable potential generated by the
superpotential (\ref{superpotestable}), $m=1$ and $c=0.24$.}}}%
\label{fig_potestable}%
\end{figure}

As an example of the preceding results consider the superpotential
\begin{equation}
W(\phi)=\frac{1}{2}m^{2}(\phi^{2}+c)\left(  e^{-\phi}+e^{\phi}\right).
\label{superpotestable}
\end{equation}
This superpotential is positive for any value of $\phi$ only if $c>0$. To it corresponds a stable scalar potential (\ref{potencial}), with a stable minimum at $\phi=0$ and $V(0)=-c^2m^4$, see Fig. \ref{fig_potestable}. In the following, in the figures we set $\kappa=1$. Thus the condition that the superpotential does not vanish, requires that this potential is negative in a neighborhood of the minimum. This class of potentials have been studied in \cite{linde1}, where it has been shown that they describe universes which after inflation stop to expand, and eventually contract again to a singularity. In particular, these potentials can be the basis for models of cyclic universes. Moreover, the evolution described by these potentials is similar to the behavior resulting from potentials unbounded from below \cite{linde1,linde2}.

\begin{figure}[h]
\centering
\includegraphics[height=3.5cm,width=4.75cm]{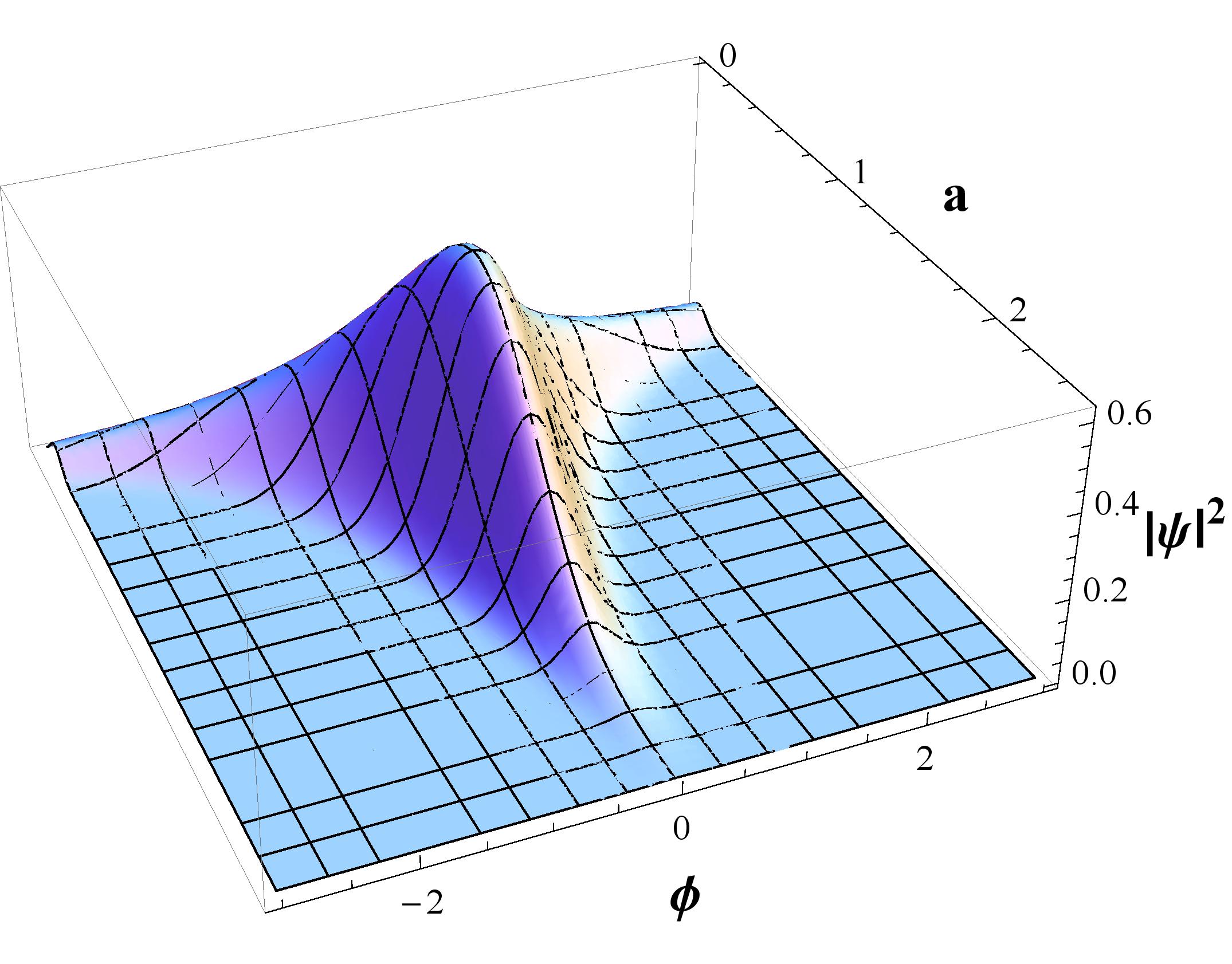}
\caption{{\protect\footnotesize {Dependence profile of $|\psi(a,\phi))|^2$ for  the superpotential (\ref{superpotestable}), $m=1$ and $c=0.24$.}}}%
\label{fig_psi1_aT_1}%
\end{figure}

\begin{figure}[h]
\centering
\includegraphics[height=3.5cm,width=4.75cm]{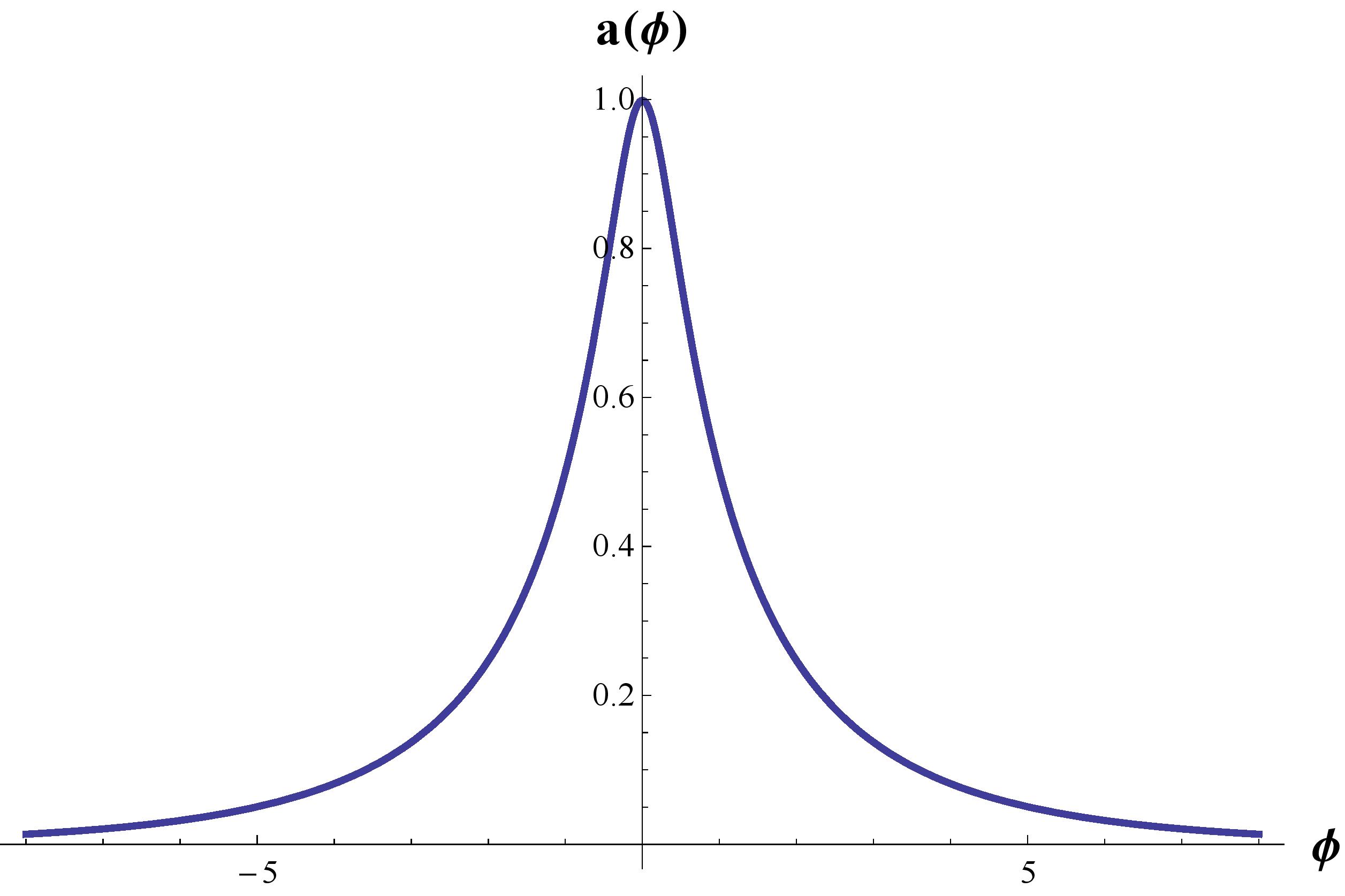}
\caption{{\protect\footnotesize {Profile of $a$ for  the superpotential (\ref{superpotestable}), $m=1$ and
$c=0.24$.}}}%
\label{fig_a1T_1}%
\end{figure}

\begin{figure}[h]
\centering
\includegraphics[height=3.5cm,width=4.75cm]{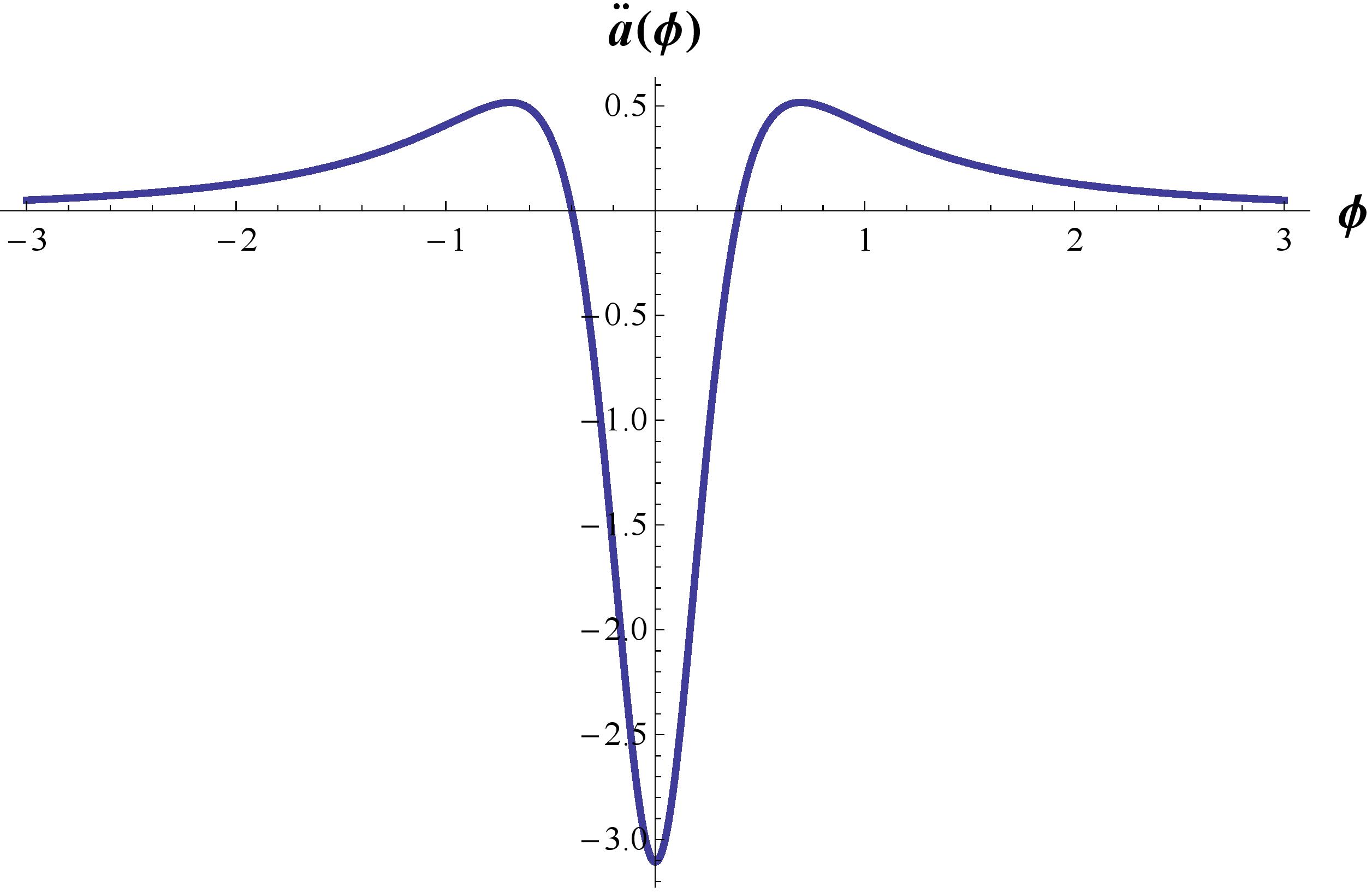}
\caption{{\protect\footnotesize {Profile of $\ddot{a}$ for  the superpotential (\ref{superpotestable}), 
$m=1$ and $c=0.24$.}}}%
\label{fig_addot_1}%
\end{figure}

\begin{figure}[h]
\centering
\includegraphics[height=3.5cm,width=4.75cm]{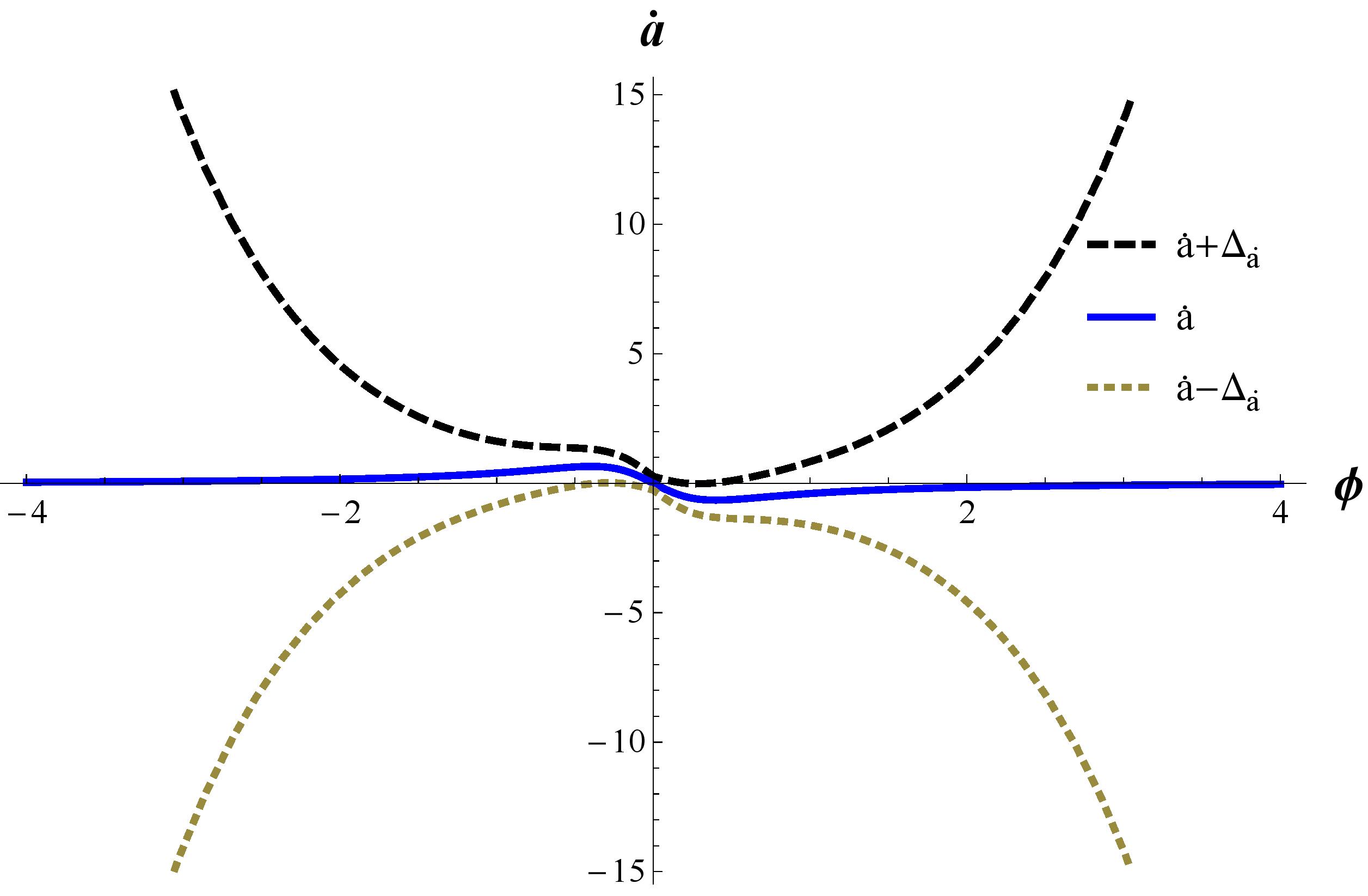}
\caption{{\protect\footnotesize {Profile of $\dot a$, including its quantum fluctuations, for the superpotential (\ref{superpotestable}), $m=1$ and
$c=0.24$.}}}%
\label{fig_fluctuaciones_1}%
\end{figure}

The wave function profile corresponding to (\ref{superpotestable}) is shown in Fig. \ref{fig_psi1_aT_1}.
Further, the evolution of the mean value of the scalar factor, $a(\tau)$, is shown in Fig. \ref{fig_a1T_1}, where we have used the
liberty to fix the free parameter $c$ in order to produce a profile
where $a(\tau=0)=1$. This evolution is consistent with the results of \cite{linde1}.
Thus, we have a growing Universe from the past at
$\tau=-\infty$ to the time $\tau=0$, when it reaches its
maximum, and then it starts to collapse as $\tau\rightarrow\infty$.
It can be seen that this potential satisfies the usual
initial conditions $\dot{a}(\tau) \to 0$ and $\ddot{a}(\tau) \to 0$ as $\tau \to-\infty$,
and also exhibits this behavior at $\tau \to\infty$. Furthermore, we depict the
corresponding behavior for $\ddot{a}(\tau)$ in Fig.
\ref{fig_addot_1}.
In Fig. \ref{fig_fluctuaciones_1} we show the quantum fluctuations of the velocity of the scale factor, which correspond to the fluctuations of the red shift. As mentioned, these fluctuations reduce in the region corresponding to the present era.

Finally, for the Hubble parameter we get
\begin{equation}
H(\tau)=\frac{c+(\tau-2)\tau-[c+(\tau+2)\tau]e^{2\tau}}{3\left(c+\tau^{2}\right) \left(1+e^{2\tau}\right)}.
\end{equation}
It has limits $H\rightarrow 1/3$ as $\tau\rightarrow-\infty$, and $H\rightarrow -1/3$ as $\tau\rightarrow\infty$.
\subsection{Unstable potential}\label{upot}

\begin{figure}[h]
\centering
\includegraphics[height=3.5cm,width=4.75cm]{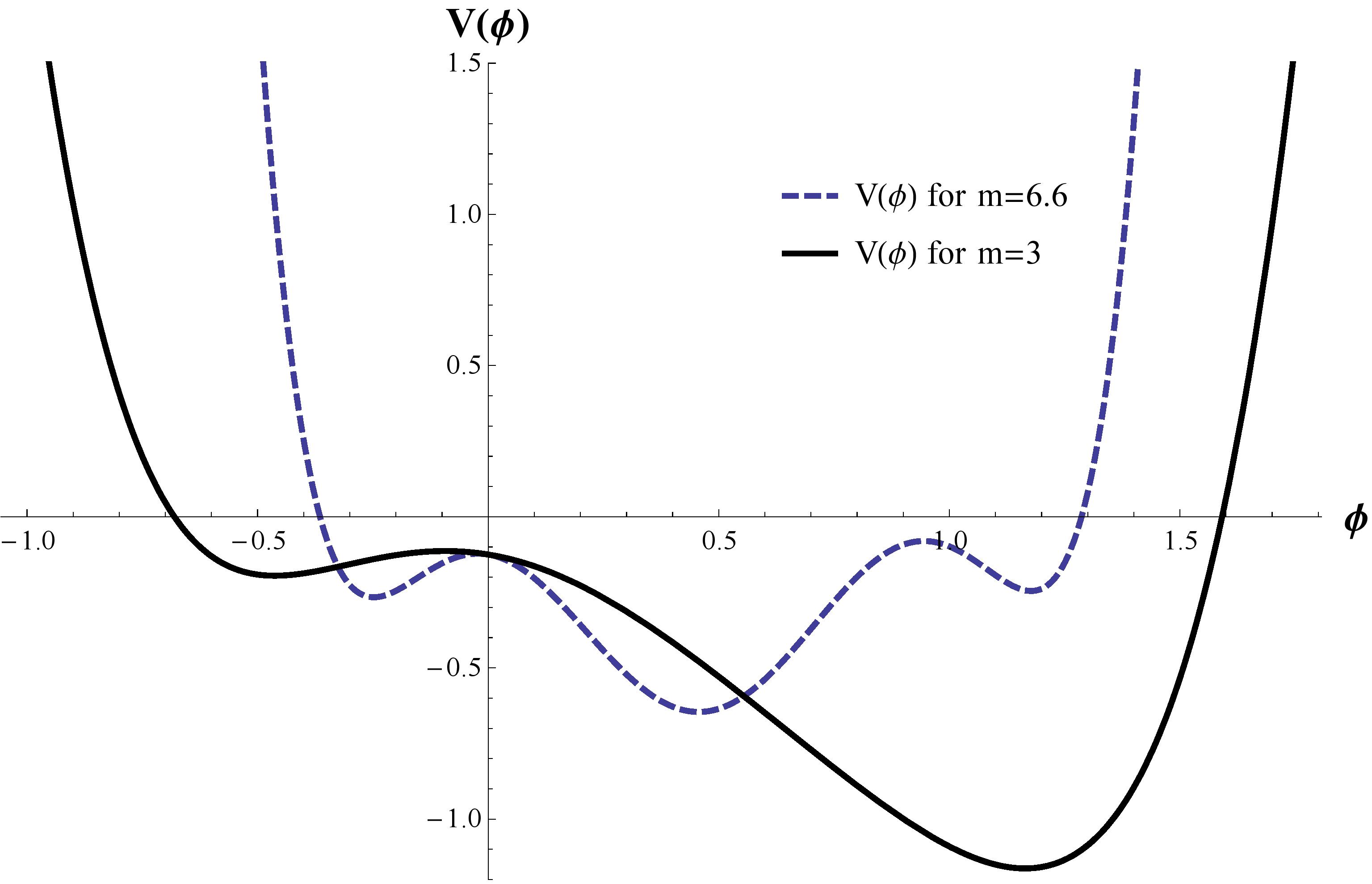}
\caption{{\protect\footnotesize {Central region of the unstable potential generated by the
superpotential (\ref{superpotprevio}), $m=3$ for the solid line and $m=6.6$ for the dashed line.}}}%
\label{fig_potinestable}%
\end{figure}

\begin{figure}[h]
\centering
\includegraphics[height=3.5cm,width=4.75cm]{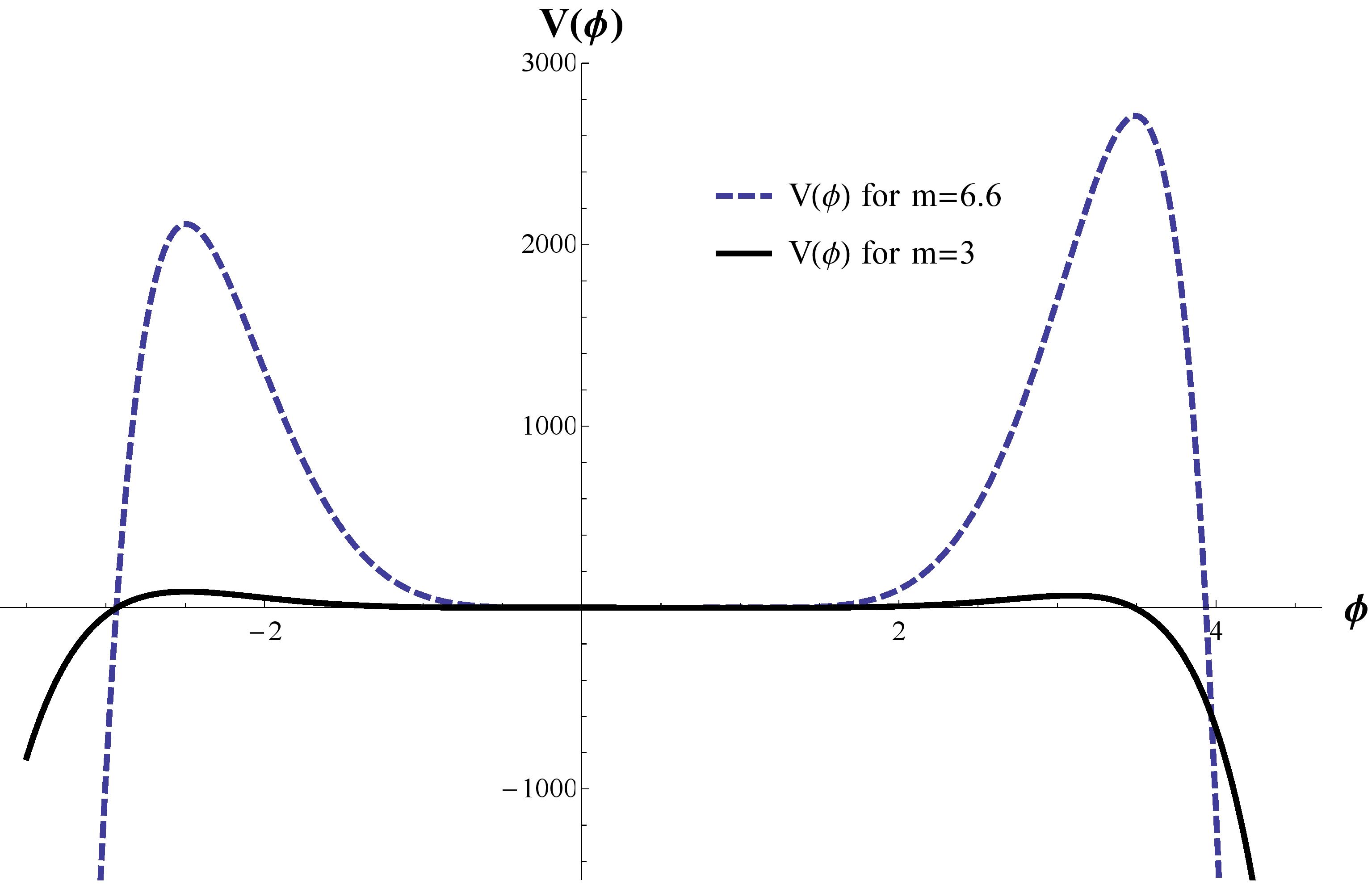}
\caption{{\protect\footnotesize {Unstable potential generated by the
superpotential (\ref{superpotprevio}), $m=3$ for the solid line and $m=6.6$ for the dashed line. The details of its central region are shown in Fig. \ref{fig_potinestable}.}}}%
\label{fig_potinestable_inf}%
\end{figure}

\begin{figure}[h]
\centering
\includegraphics[height=3.5cm,width=4.75cm]{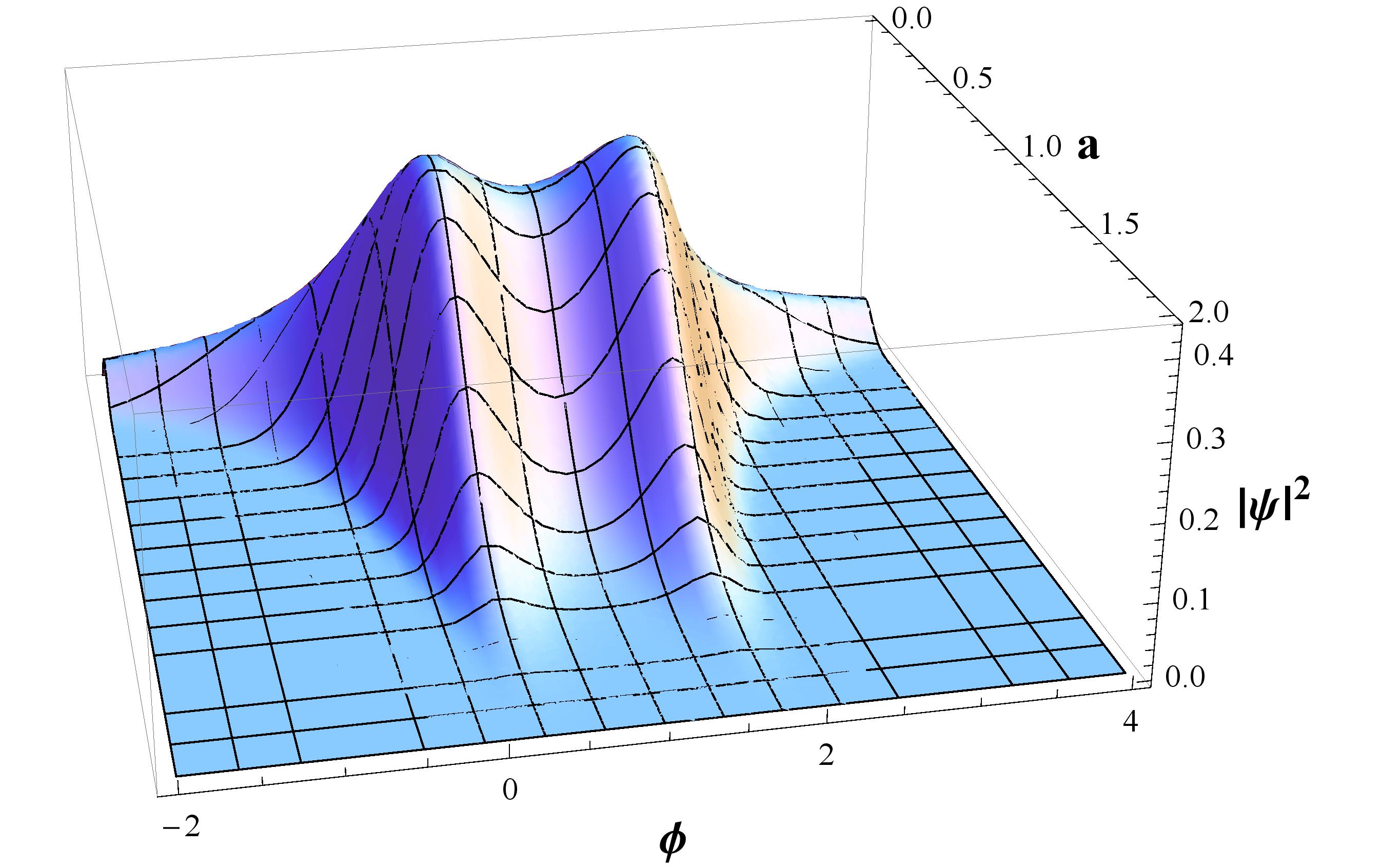}
\caption{{\protect\footnotesize {Dependence profile of $|\psi(a,\phi)|^2$ for the
superpotential (\ref{superpotprevio}), $m=6.6$.}}}%
\label{fig_psi1_aT_2}%
\end{figure}

\begin{figure}[h]
\centering
\includegraphics[height=3.5cm,width=4.75cm]{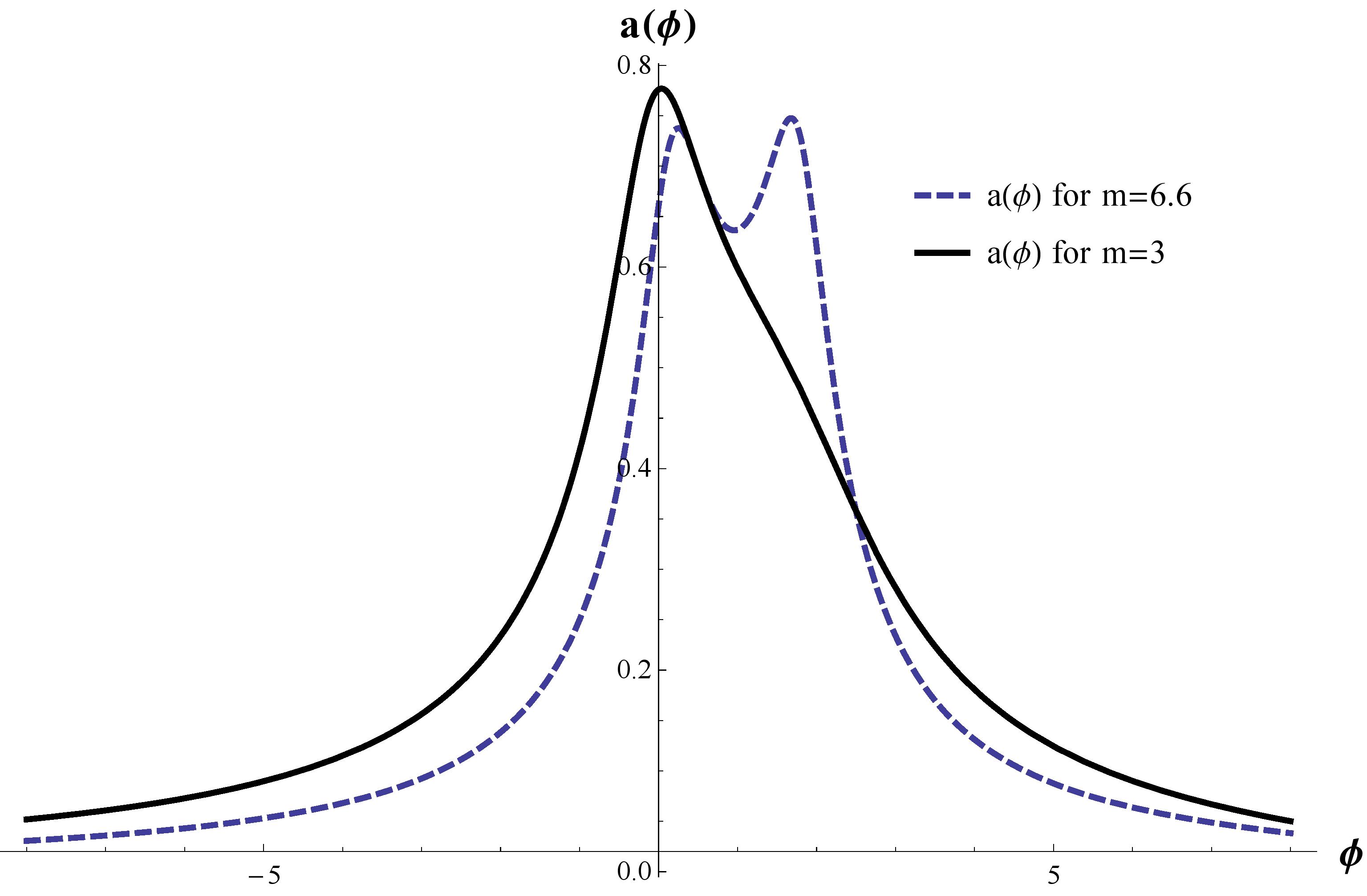}
\caption{{\protect\footnotesize {Profile of $a$ for the
superpotential (\ref{superpotprevio}), $m=3$ for the solid line and $m=6.6$ for the dashed line.}}}%
\label{fig_a1T_2}%
\end{figure}

\begin{figure}[h]
\centering
\includegraphics[height=3.5cm,width=4.75cm]{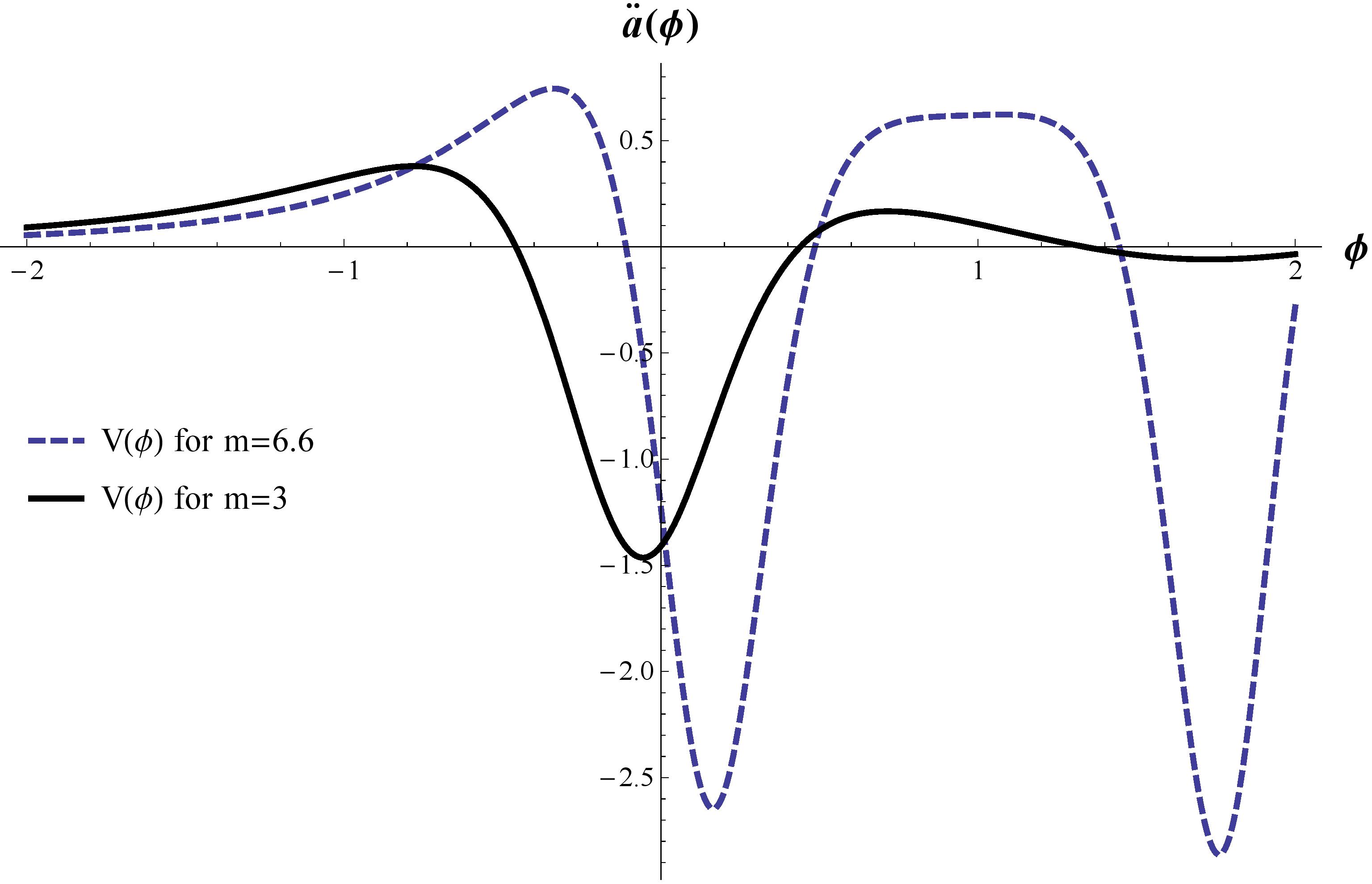}
\caption{{\protect\footnotesize {Profile of $\ddot{a}$ for the
superpotential (\ref{superpotprevio}), $m=3$ for the solid line and $m=6.6$ for the dashed line.}}}%
\label{fig_addot_2}%
\end{figure}

\begin{figure}[h]
\centering
\includegraphics[height=3.5cm,width=4.75cm]{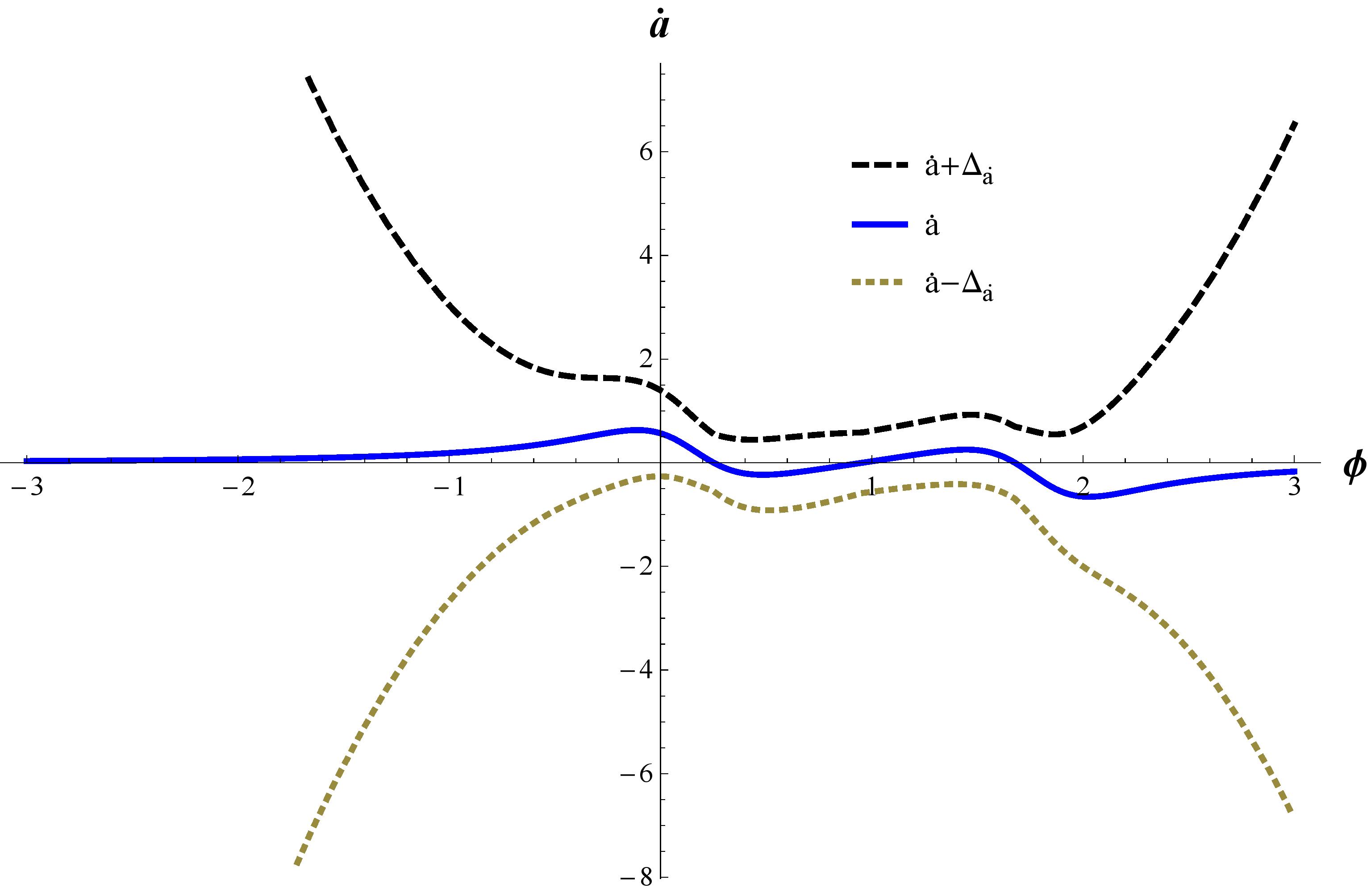}
\caption{{\protect\footnotesize {Profile of $\dot a$, including its quantum fluctuations, for the
superpotential (\ref{superpotprevio}), $m=6.6$.}}}%
\label{fig_fluctuaciones_2}%
\end{figure}

Another example is the superpotential
\begin{equation}
W(\phi)=\frac{1}{\sqrt{2}}\left(-\frac{1}{12}m^{2}\phi^{3}+\frac{1}{24}m^{2}\phi^{4}+e^{\phi}\right), \label{superpotprevio}
\end{equation}
which is positive for any value of $m$.
In \cite{previo} it has been shown that the superpotential corresponding by (\ref{potencial}) to a quartic tachyonic potential, is given by an infinite power series, which is locally approximated by (\ref{superpotprevio}). However, (\ref{superpotprevio}) is interesting by itself, because its scalar potential $V(\phi)$ has properties which lead to instabilities, as discussed in \cite{linde1}. Indeed, $V(\phi)$ has negative local minima, being otherwise positive in a neighborhood of the origin, and at the same time it is unbounded from below for $\phi\rightarrow\pm\infty$. This behavior is shown for $m=3$ and $m=6.6$ as follows. In Figure \ref{fig_potinestable} are shown the details of the local minima that are located in the central part of Figure  \ref{fig_potinestable_inf}, the last corresponding to the large scale behavior.  From the last Figure, it is interesting to note that $-V(\phi)$ has two negative minima, including the global one; hence it corresponds to a potential similar to the one of the example of the preceding Subsection. 
For $m=3$, the graphics of the wave function squared is quite similar to Fig. \ref{fig_psi1_aT_1}, and for $m=6.6$ it is shown in Fig. \ref{fig_psi1_aT_2}. Further, the mean value of the scale factor is shown in Fig. \ref{fig_a1T_2}. 
Its evolution is consistent with the one of the previous example, Fig. \ref{fig_a1T_1}. For both values of $m$ there is a singularity at the origin and at some point a collapse; for $m=3$ the evolution is like in Fig. \ref{fig_a1T_1}, but for $m=6.6$ there is a second, shorter inflationary phase before collapse. Note that this last structure cannot be associated directly to the structure of the potential for this value of $m$ in Fig. \ref{fig_potinestable}, as can be seen from a comparison of the details of both Figures.
It can be easily seen that this model satisfies as well the boundary conditions $\dot{a}(t)\to0$ and 
$\ddot{a}(t) \to 0$, as $t \to\pm\infty$. The acceleration $\ddot{a}(t)$ is shown in Figure
\ref{fig_addot_2}.
In Fig. \ref{fig_fluctuaciones_2} the quantum fluctuations of the velocity of the scale factor are shown for $m=6.6$. Similar to the preceding case, these fluctuations reduce notably in the region corresponding to the present era.

The Hubble parameter (\ref{Hubble_T}) is in this case
\begin{equation}
H(\tau)=\frac{2 m^{2} (3-2 \tau) \tau^{2}-24 e^{\tau}}{3 m^{2} (\tau-2) \tau^{3}+72 e^{\tau}}.
\end{equation}
It has the limits $H\rightarrow 0$ as $\tau\rightarrow-\infty$, and $H\rightarrow -1/3$ as $\tau\rightarrow\infty$.

\section{Conclusions}

We have studied the worldline supersymmetric theory of FRW universe, with a
scalar field. The action is constructed in the superfield formalism following \cite{previo} and \cite{cupa}. The quantization is formulated in the canonical formalism. The operator ordering ambiguities in the supersymmetric constraints are solved by means of the Weyl ordering, leading to ``zero point" contributions. The Hilbert space has an indefinite inner product. However, provided the superpotential is nowhere vanishing, there is only one consistent solution to the constraint equations. This solution is bosonic and can be chosen to have positive norm and can be normalized. The examples considered, which correspond to stable and unstable potentials, with probability densities shown in Figures \ref{fig_psi1_aT_1} and \ref{fig_psi1_aT_2}, suggest an interpretation of the scalar field $\phi$ as clock, in such a way that a time dependent, conditional probability density is obtained from the section of constant $\phi$ of the wave function \cite{Kuchar}, properly normalized. Considering that the actual universe is classical, from this probability density the measurable information is obtained via mean values of the observables, in our case limited to the scale factor. With this setting, the Heisenberg uncertainty relation for the scale factor and its canonical momentum is satisfied.
The two considered examples, whose corresponding scalar potentials are shown in Figures \ref{fig_potestable},  \ref{fig_potinestable} and \ref{fig_potinestable_inf}, are consistent with the results of references \citep{linde2,linde1}, as they lead to inflationary scenarios with initial and final singularities. Moreover, there may be more than one inflationary phases, as happens in the example of Subsection \ref{upot} for $m=6.6$. The quantum fluctuations of the velocity $\dot a$, corresponding to the actual measurable quantity regarding the scale factor, the red shift, are computed from the fluctuations of $\pi_a$, and are shown in Figures \ref{fig_fluctuaciones_1} and \ref{fig_fluctuaciones_2} for the worked examples. It is remarkable that these fluctuations reduce considerably their size in the region corresponding to the present era. It would be interesting to make a more detailed study of more realistic cosmological models, as well as under the inclusion of additional matter fields. It could be explored also the introduction of effects like noncommutativity \cite{nc}.

\vskip 1truecm \centerline{\bf Acknowledgments} We thank VIEP-BUAP and
PROFOCIE-SEP for the support.



\vskip 2truecm


\end{document}